\documentclass[runningheads]{llncs}
\usepackage[T1]{fontenc}
\usepackage{svg}
\usepackage{array}
\usepackage{tabularx}
\usepackage{amsmath}
\usepackage{subcaption}
\usepackage{graphicx}
\usepackage{booktabs}
\usepackage{multirow}
\usepackage{multicol}
\usepackage{xspace}
\usepackage{xcolor}
\usepackage{makecell}

\usepackage[table,xcdraw]{xcolor}

\begin{document}
\title{A Distributed Framework for Causal Modeling of Performance Variability in GPU Traces}
\titlerunning{Causal Modeling of Performance Variability}

\author{
Ankur Lahiry\orcidID{0009-0002-6770-4320}\inst{1} \and
Ayush Pokharel\orcidID{0009-0004-1254-9436}\inst{1} \and
Banooqa Banday\orcidID{0009-0001-5984-393X}\inst{1} \and
Seth Ockerman\orcidID{0000-0002-5692-9359}\inst{2} \and
Amal Gueroudji\orcidID{0009-0004-4830-3139}\inst{3} \and
Mohammad Zaeed\orcidID{0000-0002-3312-5462}\inst{1} \and
Tanzima Z. Islam\orcidID{0000-0003-2877-5871}\inst{1} \and
Line Pouchard\orcidID{0000-0002-2120-6521}\inst{4}
}

\authorrunning{A. Lahiry et al.}

\institute{Texas State University, San Marcos, TX 78666, USA\\
\email{\{vty8, ssu22, banooqa, tanzima\}@txstate.edu, tz2201@gmail.com} \and
University of Wisconsin-Madison, Madison, WI 53706, USA\\
\email{sockerman@cs.wisc.edu} \and
Argonne National Laboratory, Lemont, IL 60439, USA\\
\email{agueroudji@anl.gov} \and
Sandia National Laboratories, Albuquerque, NM 87123, USA\\
\email{lcpouch@sandia.gov}
}

\maketitle              % typeset the header of the contribution
\begin{abstract}
Large-scale GPU traces play a critical role in identifying performance bottlenecks within heterogeneous High-Performance Computing (HPC) architectures. However, the sheer volume and complexity of a single trace of data make performance analysis both computationally expensive and time-consuming. To address this challenge, we present an end-to-end parallel performance analysis framework designed to handle multiple large-scale GPU traces efficiently. Our proposed framework partitions and processes trace data concurrently and employs causal graph methods and parallel coordinating chart to expose performance variability and dependencies across execution flows. Experimental results demonstrate a 67\% improvement in terms of scalability, highlighting the effectiveness of our pipeline for analyzing multiple traces independently. 

\end{abstract}

\section{Introduction}
% Modern High-Performance Computing (HPC) systems increasingly rely on heterogeneous architectures consisting of multi-core CPUs, GPUs, advanced memory hierarchies, high-speed networks, and multi-NIC nodes. GPUs, in particular, are widely used due to their inherent high parallelism and efficiency in accelerating compute-intensive applications, such as various large-scale scientific simulations and AI training and inference. However, despite the potential for performance excellence, the increasing complexity of heterogeneous systems introduces substantial challenges in performance and optimization. Even minor inefficiencies in the kernel-level or memory-level can result in significant performance bottlenecks, causing drastic degradation of the system's performance. To provide a seamless experience for the users, analyzing the GPU traces is crucial. 
% GPU traces provide detailed characterizations of execution behavior and insights into the underlying causes of performance inefficiencies.
% However, analyzing large-scale GPU traces is crucial due to the sheer size of the data, which often reaches hundreds of gigabytes or more per execution. Sequential analysis is computationally expensive and time-consuming. 
Modern high-performance computing (HPC) systems increasingly integrate heterogeneous architectures combining multi-core CPUs, GPUs, advanced memory hierarchies, and high-speed interconnects. GPUs, in particular, drive most compute-intensive workloads due to their massive parallelism and efficiency in accelerating large-scale simulations and AI models.
Yet, this architectural complexity introduces serious optimization challenges. Even small inefficiencies at the kernel or memory level can create performance bottlenecks, leading to severe throughput degradation. Understanding these inefficiencies requires detailed GPU trace analysis—an indispensable tool for characterizing execution behavior and uncovering root causes of performance loss.
However, large-scale GPU traces can reach hundreds of gigabytes per run, making sequential analysis prohibitively slow and resource-intensive.
To address this challenge, we develop an end-to-end parallel performance analysis framework for large-scale GPU traces. Our work makes three key contributions:
% In this work, we propose an end-to-end parallel performance analysis framework designed to efficiently process multiple large-scale GPU traces. \begin{itemize}
%     \item A distributed framework that efficiently partitions large trace datasets across multiple compute nodes, enabling the concurrent analysis of different trace segments and significantly reducing overall processing time.
%     \item We introduce novel algorithms that identify performance variability patterns across multiple execution traces using causal relationships, providing deeper insights into the root causes of performance degradation in heterogeneous HPC systems.
%     \item  We present a complete end-to-end pipeline that seamlessly handles the entire workflow from trace ingestion to performance insight generation, making large-scale GPU performance analysis accessible and efficient for HPC practitioners.
% \end{itemize}
\vspace{-0.5em}
\begin{itemize}
    \item Design a scalable distributed framework that parallelizes GPU trace analysis across compute nodes, achieving high throughput and substantially reducing time-to-insight.
    \item Leverage causal inference–based algorithms that extract variability patterns and link them to system-level inefficiencies, revealing root causes of performance degradation across executions.
    \item Integrate a complete end-to-end pipeline that automates data ingestion, parallel processing, and insight generation, turning raw GPU telemetry into actionable performance knowledge.
\end{itemize}

% Together, these contributions establish a unified, scalable framework that redefines how large-scale GPU traces are processed and interpreted. The distributed design ensures efficient utilization of compute resources, while the causal algorithms uncover hidden dependencies that traditional profiling overlooks. The integrated workflow eliminates manual bottlenecks, enabling reproducible, high-fidelity analysis at scale. Collectively, the framework bridges the gap between low-level trace data and high-level performance understanding—empowering HPC practitioners to diagnose, optimize, and sustain efficiency across heterogeneous systems.

%\section{Background}
%\input{Section/background}

\section{Methodology}

\subsection{Features}

\begin{table}[!t]
\centering
\caption{Profiler column names are simplified and grouped into four categories to enhance clarity. The original columns remain in the dataset, but the simplified names improve visualization and interpretation in the causal graphs presented later.}
\label{tab:profiler_grouped}
\resizebox{\textwidth}{!}{
\begin{tabular}{|p{0.6\linewidth}|p{0.4\linewidth}|}
\hline
\multicolumn{2}{|c|}{\cellcolor{green!25}\textbf{Identification and Context}} \\ \hline
\textbf{Original} & \textbf{Simplified} \\ \hline
Kernel\_Name, Device, CC, Process\_Name, Host\_Name & Kernel\_Name, Device, CC, Process\_Name, Host\_Name \\ \hline

\multicolumn{2}{|c|}{\cellcolor{orange!25}\textbf{Kernel Launch Configuration}} \\ \hline
Grid\_Size, Workgroup\_Size & grid\_size, workgroup\_size \\ \hline
LDS\_Per\_Workgroup, Scratch\_Per\_Workitem & lds\_per\_workgroup, scratch\_per\_workitem \\ \hline
{[}Arch, Accum{]} VGPR, SGPR, wave\_size & {[}arch, accum{]}\_vgpr, sgpr, wave\_size \\ \hline

\multicolumn{2}{|c|}{\cellcolor{blue!20}\textbf{Temporal Characteristics}} \\ \hline
runtime & runtime \\ \hline
SQ\_[WAVES, BUSY\_CYCLES, CYCLES] & sq\_[waves, busy\_cycles, cycles] \\ \hline
SQ\_INSTS\_[VALU, SALU, VMEM] & sq\_insts\_[valu, salu, vmem] \\ \hline

\multicolumn{2}{|c|}{\cellcolor{purple!20}\textbf{Memory System Activity}} \\ \hline
TCC\_[RW\_REQ, BUSY, TAG\_STALL, HIT, MISS, READ, WRITE]\_sum & tcc\_[rw\_req, busy, tag\_stall, hit, miss, read, write] \\ \hline
TCP\_PENDING\_STALL\_CYCLES\_sum & tcp\_pending\_stall \\ \hline
TA\_BUFFER\_[READ, WRITE]\_WAVEFRONTS\_sum & ta\_buf\_[read, write] \\ \hline
TD\_[LOAD, STORE]\_WAVEFRONT\_sum & td\_[load, store] \\ \hline
\end{tabular}
}
\end{table}
In GPU computing, a kernel is a user-defined function executed in parallel by thousands of threads. Each kernel launch specifies a grid–block configuration that defines how threads are organized and scheduled. During execution, kernels interact with multiple memory hierarchies—registers, shared, global, and cache—each affecting latency and throughput. Although a kernel’s internal logic remains constant across invocations, its performance metrics such as runtime and memory transfer time often fluctuate due to resource contention and hardware variability. Identifying this variability requires analyzing the feature space captured during profiling. GPU profiling tools such as NVIDIA CUPTI~\cite{2016-6-1249}, Klaraptor~\cite{brandt2019klaraptor}, and ROCm~\cite{shafie2021designing} expose different sets of kernel features. Table~\ref{tab:profiler_grouped} summarizes the grouped mapping of these profiler columns to their simplified names used for subsequent causal analysis.
% \begin{itemize}
%   \item \textbf{Identification and Context}:  
%   Unique descriptors of the kernel launch, such as dispatch or stream IDs, kernel names, GPU and process/thread identifiers, and queue information. These attributes tie each measurement back to the originating workflow and hardware context.
%   \item \textbf{Kernel Launch Configuration}:  
%   Static properties of the launch, including grid and block dimensions, workgroup sizes, wavefront/warp sizes, and register and shared memory allocations per thread or per workgroup. These features reflect how much parallelism and on-chip resources a kernel instance requests.
%   \item \textbf{Temporal Characteristics}:  
%   Timestamps and durations that capture when a kernel is dispatched, starts, and completes. These fields are used to derive execution time, overlap with other operations, and latency distributions.
%   \item \textbf{Memory System Activity}:  
%   Aggregated counters for cache and memory hierarchy behavior, including read/write/atomic requests, hits and misses, stalls, credit back-pressure, and evictions across different cache levels (e.g., TCC, SQC, TCP). These features quantify memory throughput, efficiency, and contention.
% \end{itemize}

\subsection{Targets}
\subsubsection{Kernel Variability}
Let $K$ be the set of all kernel identifiers in a given performance dataset. For a given kernel $k \in K$, which is repeated $n$ times, we define $\mathcal{R}_k = \{r_{k,1}, r_{k,2}, \dots, r_{k,n_k}\}$ as the execution of that kernel. We define kernel variability as the extent to which a kernel execution time differs from the \textbf{average kernel execution time for each configuration} calculated by $
\bar{r}_k = \frac{1}{n_k} \sum_{i=1}^{n_k} m\bigl(r_{k,i}\bigr)$.
So, the \textbf{run-to-run variation} of a given kernel can be calculated by measuring the difference between the execution time of each repeated run and the average kernel execution time for that configuration: $
v_{\text{abs}}\bigl(r_{k,i}\bigr) = 
\bigl|\, m\bigl(r_{k,i}\bigr) - \bar{m}_k \,\bigr|$.

\subsubsection{Memory Stall Variablity}
GPU kernels often overlap computation and memory operations. However, in many cases, memory operations extend beyond the kernel's active execution window, creating stalls. To quantify this behavior, we track the start and end times of both the kernel execution and the corresponding memory transfers. For each kernel \(k \in K\), we observe a set of runs \(\mathcal{R}_k = \{(r_{k,i}^{s}, r_{k,i}^{e}) \mid i=1,\dots,n_k\}\), where \(r_{k,i}^{s}\) and \(r_{k,i}^{e}\) denote the start and end times of the \(i\)-th run of kernel \(k\), respectively. Similarly, we record the associated memory transfer intervals \(\mathcal{M}_k = \{(m_{k,i}^{s}, m_{k,i}^{e}) \mid i=1,\dots,n_k\}\), where \(m_{k,i}^{s}\) and \(m_{k,i}^{e}\) denote the start and end times of the memory transfer linked to the \(i\)-th run. We compute the absolute difference between kernel duration and the memory transfer interval and define the difference as memory stall variability.

% We then derive three quantities for each run \(r_{k,i}\):

% \begin{itemize}
%     \item \textbf{Kernel execution interval} — the duration of kernel activity: \tzi{you need to give the physical significance of the equation before the equation, introduce every term of the equation, and give physical significance of each of the subparts of an equation if you want to use equation. do not use trivial equations, not necessary. It makes a paper unreadable.}
%     \[
%     m_{\text{kern}}(r_{k,i}) = \text{end}(r_{k,i}) - \text{start}(r_{k,i}).
%     \]

%     \item \textbf{Memory operation interval} — the duration of memory activity: \tzi{remove the equation and give physical significance of the variable. Are you even using it in the result section? if not, remove it altogether.}
%     \[
%     m_{\text{mem}}(r_{k,i}) = \text{mem\_end}(r_{k,i}) - \text{mem\_start}(r_{k,i}).
%     \]

%     \item \textbf{Memory stall duration} — the non-negative \tzi{excess of memory -- what does excess of memory mean?????????????????????} activity beyond kernel execution:
%     \[
%     d_{\text{mem}}(r_{k,i}) =
%     \max\!\Bigl(0,\,
%     m_{\text{mem}}(r_{k,i}) - m_{\text{kern}}(r_{k,i})\Bigr).
%     \]
% \end{itemize}

\subsection{Binning to Reduce Samples}
\label{subsection:binning}

GPU traces record detailed information about every kernel execution over time and often reach massive sizes, making full-trace analysis both memory-intensive and difficult to interpret. To manage this complexity, the trace is divided into fixed-length time segments, called bins, where each bin represents a portion of the GPU's execution timeline. All kernel activities that begin within a given bin are grouped together, and a set of summary statistics—such as minimum, maximum, mean, variance, standard deviation, quartiles, and sample count—is computed to describe their behavior. These statistics quantify how stable or variable the performance is within that time window: for instance, variance and standard deviation measure how widely the data values differ, while quartiles help detect unusual outliers.
By analyzing the trace one bin at a time, regions showing significant runtime or resource usage fluctuations can be isolated. Bins are then ranked according to their variability, allowing the most critical time periods to be examined in greater depth. This binning approach reduces data size, preserves essential performance trends, and enables independent analysis across different parts of the execution timeline.

\subsection{Distributed Parallel Architecture}

\begin{figure}[!hbtp]
\centering\includegraphics[width=0.9\linewidth]{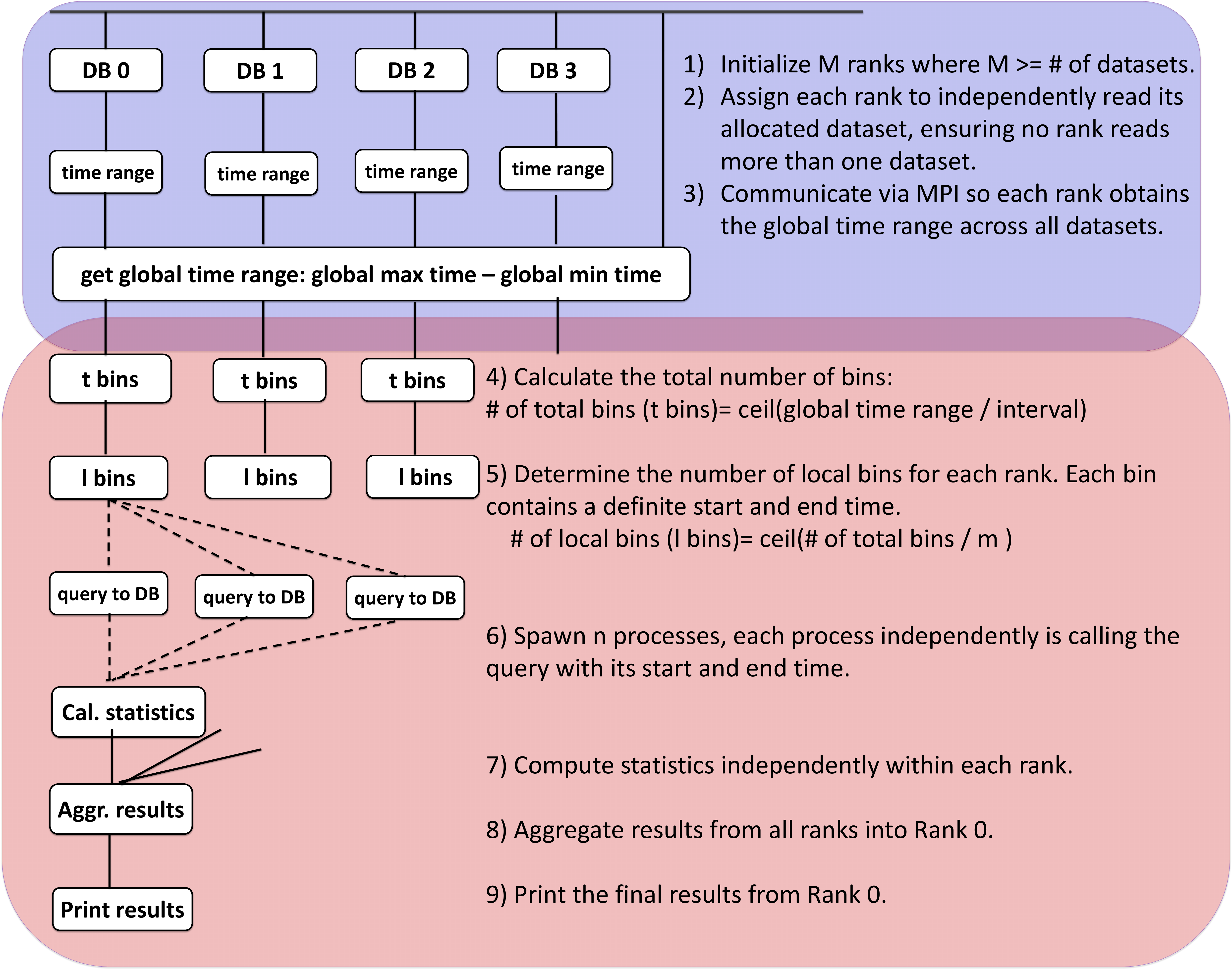}
    \caption{Overview of the parallel GPU trace analysis framework. The system combines MPI-based distributed parallelism with shared-memory threading to process large GPU activity traces efficiently. MPI ranks divide the workload across nodes (Steps 1–4), each spawning local worker processes to analyze bins in parallel (Steps 5–7). Results are then aggregated (Step 8), and high-variability bins are identified for detailed analysis (Step 9).}
    \label{fig:parallel}
\end{figure}

\vspace{-.8em}

Figure~\ref{fig:parallel} shows the overall architecture of our framework. Our system handles large numbers of GPU activity traces in parallel using hybrid MPI and shared-memory parallelism for maximum resource utilization on modern HPC systems. The program launches multiple MPI ranks, each responsible for a portion of the workload \textbf{(Step 1)}. The workload consists of bins obtained by segmenting the global execution timeline into fixed-length intervals \textbf{(Step 2)}. Ranks collaborate to compute the global time range by querying their dataset for minimum and maximum timestamps, then combining values using collective communication \textbf{(Step 3)}. Each rank is assigned a balanced subset of bins \textbf{(Step 4)}. After obtaining the end-to-end execution time, we launch worker pools, where each worker processes their assigned bins \textbf{(Step 5)}. A worker is an independent process created using Python’s multiprocessing module within an MPI rank. Each worker concurrently retrieves its assigned bins from the dataset, performs the analysis, and applies the variability-based filter to retain only the most relevant bins. Low-variability bins are discarded from further analysis, keeping only the top $K=5$ highest-variance bins for future computation like identifying the causality, memory stall etc. This hybrid approach achieves two goals: 1) parallel processing of multiple datasets and 2) scalable processing of large datasets.

% \subsection{Binning to Reduce Samples}
% \label{binning}

% After finding out the global execution timeline across all the databases, the system segments the timeline into uniform time windows. We refer to each window as a bin. Each bin corresponds to a user defined fixed length. For every database and for every bin, the program extracts all GPU kernel records whose start time falls within the interval. Within each bin, the program computes a comprehensive set of descriptive statistics: \textit{minimum}, \textit{maximum}, \textit{mean}, \textit{variance}, \textit{standard deviation}, \textit{first quartile}, \textit{third quartile} , and \textit{the number of samples} in each bin. The \textit{variance} and \textit{standard deviation} measure how widely the values in a bin are spread out overall. However, these measures can be strongly affected by extreme values, so we also calculate the \textit{quartiles} and \textit{interquartile range}. 

% Once all bins have been processed and their statistics gathered across all ranks, the system performs sorting steps. For each target metric, the bins are compared based on their variability and then sorted in descending order of variance. Finally, the program identifies the top K bins , where the default K = 5, for each feature and reports them as the most significant time windows for further analysis. 

\subsection{Causal Modeling}
Causal modeling~\cite{pearl2009causality} is a systematic approach for discovering and quantifying cause-and-effect relationships within data. Unlike correlational methods, which reveal associations without explaining why they occur, causal modeling enables reasoning about how changes in one variable may directly influence another. This shift from describing patterns to explaining mechanisms allows researchers to move from observation to intervention and prediction. A causal model encodes assumptions about how variables interact and uses these assumptions to estimate directional effects. It is typically represented as a directed graph or a system of equations that specifies which variables act as causes, which as effects, and which serve as confounders—factors that influence both. Once this structure is defined, statistical and computational techniques are applied to estimate causal strengths and to simulate counterfactuals--scenarios that ask, what would happen if a specific change or intervention were made?

% An Interquartile Range (IQR) is a measure of spread in a dataset that represents the range of the middle 50\% of the data, calculated by subtracting the first quartile (Q1) from the third quartile (Q3). The IQR is particularly useful because it is not affected by outliers or extreme values in the dataset
% [https://en.wikipedia.org/wiki/Interquartile\_range]

\subsection{Output}
% GPU performance traces contain a wealth of metrics, such as execution times, memory throughput, cache hit rates, etc., that interact in a complex way. Analyzing this high-dimensional, independent data is essential for identifying performance bottlenecks, understanding variability, and optimizing GPU workloads. To address this, we provide three complimentary visualization methods: causal graphs, bar plots, and parallel coordinate plots to understand the different facets of performance variance. Together, these visualizations enable both high-level overviews and detailed insights, providing a comprehensive understanding of how various factors contribute to GPU performance behavior. 

\subsubsection{Causal Graph for understanding influence of features on variability}
% Directed causal graphs are powerful tools for identifying the relationships between different metrics captured in GPU performance traces. Unlike simple correlation plots, causal graphs can help visualize both direct and indirect effects, enabling researchers to pinpoint which factors are crucial for identifying performance bottlenecks. Our causal graph demonstrates both the positive and negative influences, with the target metric represented by a black directed arrow and a red directed arrow, respectively. The weight associated with the arrow indicates how \% of the feature causes the performance variance. 
Directed causal graphs reveal how different GPU performance metrics influence one another, distinguishing direct from indirect effects. Unlike correlation plots, they expose which factors most strongly contribute to performance bottlenecks. In our graph, black and red arrows denote positive and negative influences, respectively, with arrow weights indicating each feature’s percentage contribution to performance variance.
% \subsubsection{Bar plots for clear and quantitative comparison}

% A bar plot provides a straightforward way to quantify and compare performance metrics across different scenarios, such as GPU kernels, configurations, or workloads. Bar plots demonstrate the feature variance of the given dataset, helping to visually identify which features have a higher variance and which features may be correlated in identifying the performance variance. Their simplicity and familiarity also make bar plots effective for communicating results to broader audiences—including engineers, researchers, and non-expert stakeholders.
\vspace{-1em}
\subsubsection{Parallel Coordinate plots for High-Dimensional Trace analysis}
% Parallel coordinate plots are suitable for high-dimensional datasets, such as GPU traces, where each axis represents different performance features, and each line corresponds to a single trace. This visualization allows researchers to simultaneously identify, inspect patterns, and correlations across many metrics, enabling the identification of clusters of traces with similar behaviors or configurations that lead to anomalous performance. 
Parallel coordinate plots visualize high-dimensional GPU trace data, where each axis represents a performance feature and each line corresponds to a single trace. This view enables simultaneous comparison across metrics, helping identify patterns, correlations, and clusters of traces that exhibit similar behaviors or performance anomalies.

\subsection{Performance}
% We measure the performance of our pipeline using a scalability study. Scalability, in this context, refers to the pipeline’s ability to handle increasing workloads, larger data volumes, or more complex operations as time progresses—without a proportional increase in execution time. A well-designed pipeline should demonstrate near-linear scalability; doubling the workload or input size results in a predictable and manageable increase in processing time. As mentioned in figure~\ref {fig:parallel}, a key feature of our pipeline is that it can handle $n$ datasets simultaneously if the datasets share common features and the target objective. Each dataset is handled by an independent processing path, with no shared-state contention or synchronization barriers. This design ensures that work on one dataset does not delay or interfere with work on another, allowing the pipeline to scale horizontally across multiple dataset inputs.
We evaluate pipeline performance through a scalability study, assessing its ability to process larger workloads and datasets without a proportional rise in execution time. As shown in Figure~\ref{fig:parallel}, the framework achieves near-linear scalability by analyzing multiple GPU trace datasets concurrently when they share common features and target metrics. Each dataset follows an independent processing path with no shared-state contention, ensuring that work on one trace does not delay another. This design enables efficient horizontal scaling across datasets and sustained performance at increasing analysis loads.

\section{Experimental Setup}
\vspace{-1em}
\begin{table}[!hbtp]
\centering
\caption{Comparison of the workflow-level and performance counter datasets, showing their sample and feature sizes, number of kernels, and associated target metric.}
 \resizebox{\textwidth}{!}{%
 \begin{tabular}{|c|c|c|c|c|c|}
 \hline
 \textbf{Category} & \textbf{Dataset} & \textbf{Samples} & \textbf{Columns} & \textbf{\# Kernels} & \textbf{Target Metric} \\ \hline

 \multirow{8}{*}{{Workflow~\cite{ockerman2025pgtiscalingspatiotemporalgnns,Guerodji2024Performance}} \multirow{8}{*}} 
  & Rank 0  & $\sim837K$ & 28 & 74 & Kernel\_variability \\ \cline{2-6}
  & Rank 0  & $\sim93M$  & 35 & 74 & memory\_stall       \\ \cline{2-6}
  & Rank 1  & $\sim837K$ & 28 & 74 & Kernel\_variability \\ \cline{2-6}
  & Rank 1  & $\sim93M$  & 35 & 74 & memory\_stall       \\ \cline{2-6}
  & Rank 2  & $\sim837K$ & 28 & 74 & Kernel\_variability \\ \cline{2-6}
  & Rank 2  & $\sim93M$  & 35 & 74 & memory\_stall       \\ \cline{2-6}
  & Rank 3  & $\sim837K$ & 28 & 74 & Kernel\_variability \\ \cline{2-6}
 & Rank 3  & $\sim93M$  & 35 & 74 & memory\_stall       \\ \hline

 \multirow{3}{*}{PerfCounter ~\cite{yazdanbakhsh2016axbench,jin2023a,petersson2023sw4}} 
 & Inversek2j-CUDA & 1642 & 8006 & 1  & gpu\_\_time\_duration.avg \\ \cline{2-6}
  & Accuracy-CUDA   & 1200 & 2119 & 1  & gpu\_\_time\_duration.avg \\ \cline{2-6}
  & SW4Lite-CUDA    & 1201 & 5036 & 17 & gpu\_\_time\_duration.avg \\ \hline

 \end{tabular}%
 }

 \label{ref:dataset}
 \end{table}

\vspace{-0.5em}
\subsection{Dataset Description}

We categorize seven datasets into two groups, as summarized in Table~\ref{ref:dataset}, which lists the number of kernels and target metric for each.

% \subsubsection{Workflow dataset}: 
% This dataset captures a single training epoch of \texttt{dask-pytorch-ddp}~\footnote{\url{https://github.com/saturncloud/dask-pytorch-ddp}} running a Diffusion Convolutional Recurrent Neural Network~\cite{li2018dcrnn_traffic} on the PeMS-Bay dataset. Training used a generalized distributed index batching method~\cite{ockerman2025pgtiscalingspatiotemporalgnns}, extending standard DDP~\cite{torch_ddp} to reduce inter-worker communication. Four parallel workers, each on an NVIDIA A100 GPU, executed the run. GPU events were collected via Nsight-Compute, and Dask events (e.g., task transitions, data transfers) via Dask-Mofka~\cite{Guerodji2024Performance}.
% \subsubsection{PerfCounter}: 
% This dataset contains raw hardware counter values from three benchmarks—Inversek2j~\cite{yazdanbakhsh2016axbench}, Accuracy~\cite{jin2023a}, and SW4lite~\cite{petersson2023sw4}—each run on a single NVIDIA A100 GPU. Inversek2j executed 10,000 iterations; Accuracy ran with 192 rows, 5,000 dimensions, top\_k=10, and 100 iterations; and SW4lite used its default point-source input. All counters were collected using Nsight-Compute.

\subsection{System}
\vspace{-0.8em}
Texas Advanced Computing Center (TACC) provides supercomputing facilities to researchers for conducting simulations. We leverage the Lonestar6 computing cluster for running all our experiments. Lonestar6 consists of 560 compute nodes and 88 GPU nodes. Each compute node is comprised of 2 AMD EPYC 7763 64-core (Milan) CPUs and 256 GB of DDR4 memory. Additionally, each of the 84 GPU nodes has 3 NVIDIA A100 GPUs with 40 GB of HBM2 high-bandwidth memory.

\section{Result}
\begin{figure}[!hbtp]
    \centering
    \includegraphics[width=\linewidth]{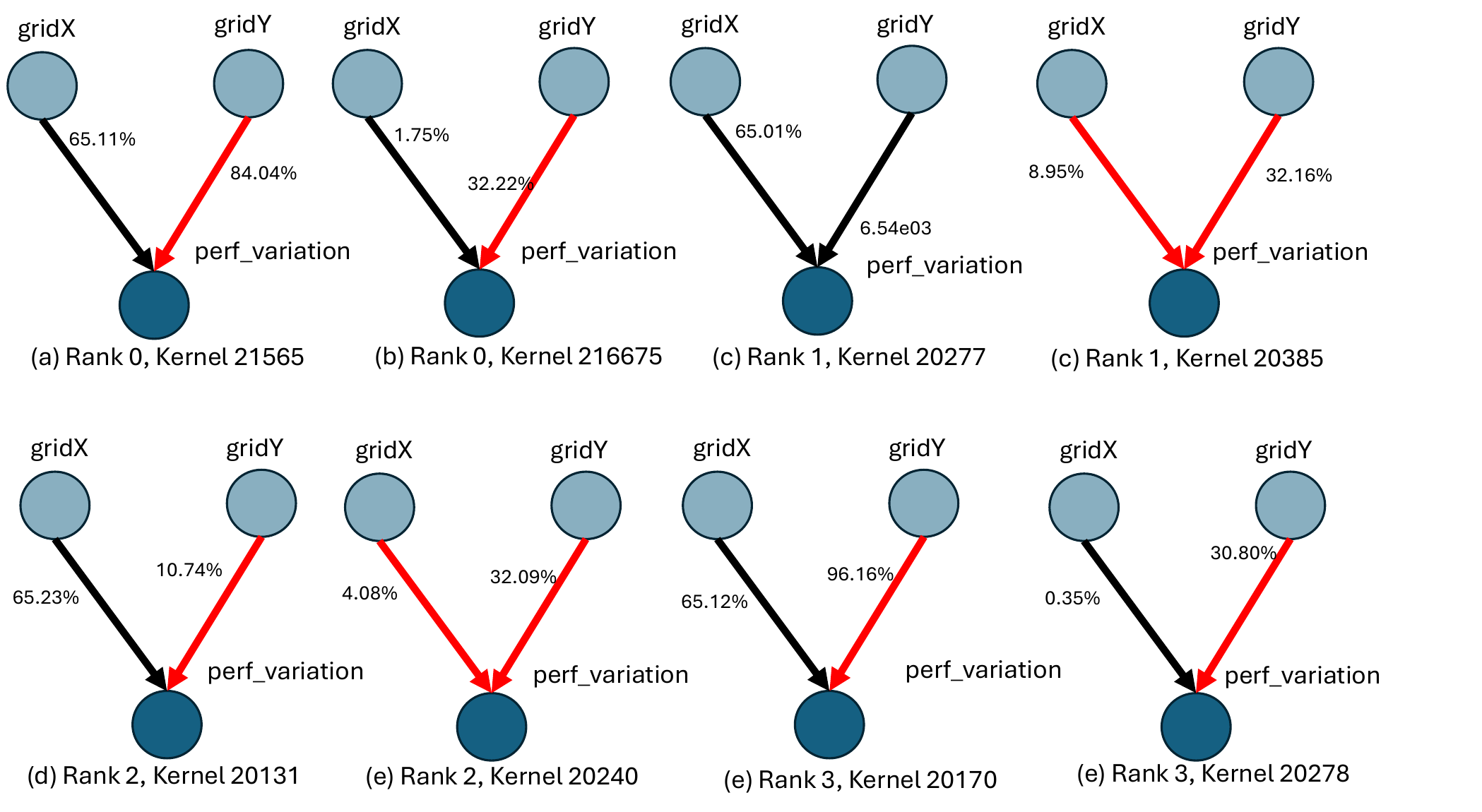}
    \caption{Kernel variability across different ranks and Kernel. Each row illustrates four examples: (a–d) correspond to Ranks 0–1 and (e–h) to Ranks 2–3. Kernel X refers to the associated \texttt{demangledName}, with two kernels randomly selected from different rank datasets. The target is \texttt{kernel\_duration}, and \texttt{perf\_variation} is computed as 
$\left| \mathrm{avg}(\texttt{kernel\_duration}) - \texttt{kernel\_duration} \right|$. Black arrows indicate positive influence (e.g., increasing \texttt{gridX} increases variability), while red arrows indicate negative influence on \texttt{perf\_variablity}. Edge weight defines change of \% of a feature has positive or negative influence on \texttt{perf\_variation}
}
    \label{fig:per_kerenl_ranks}
\end{figure}

\vspace{-1.0em}

\begin{figure}[!hbtp]
\includegraphics[width=\linewidth]{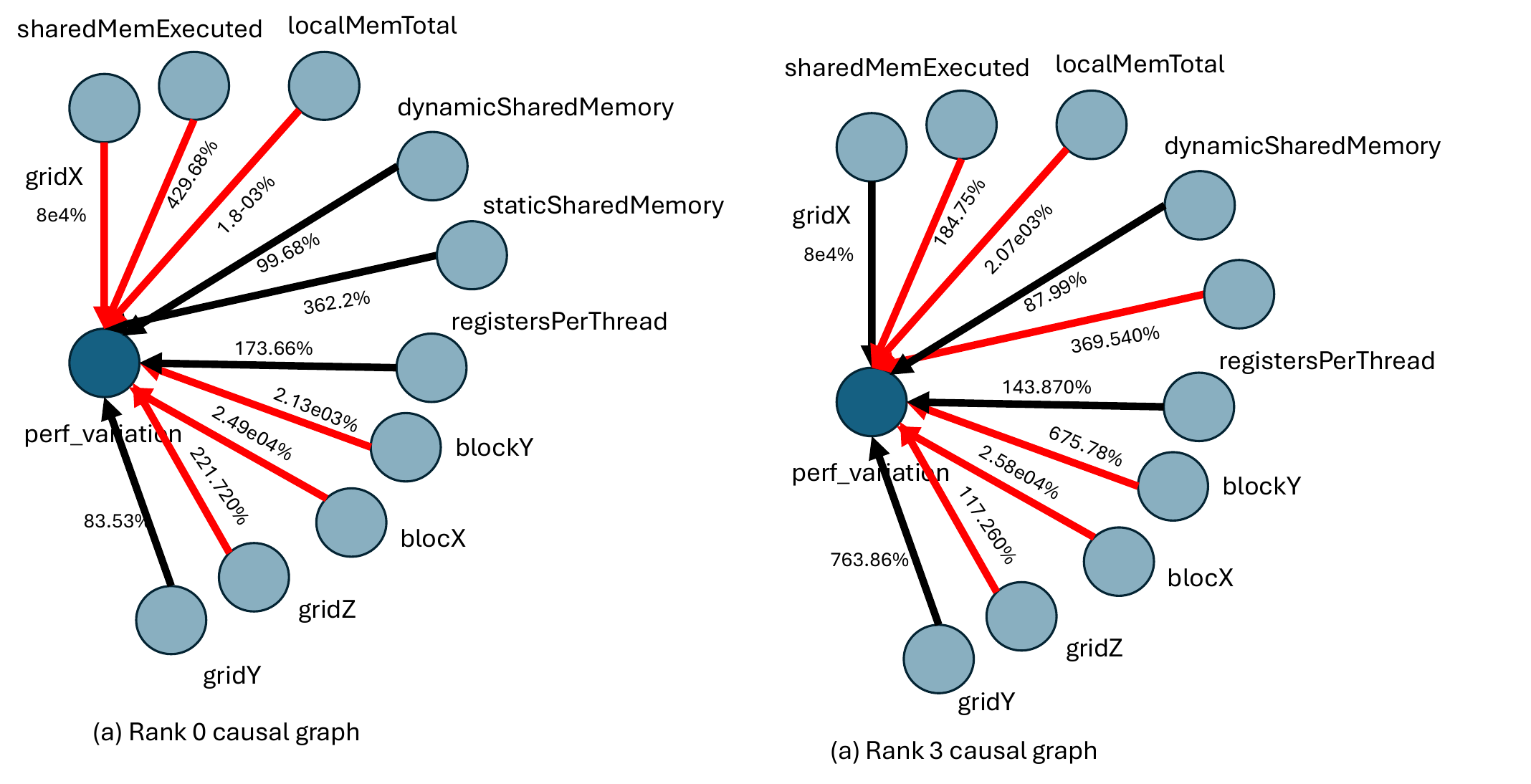}
    \caption{Overview of kernel variability across ranks and kernels.
This figure combines results from all kernels to illustrate variability across different ranks. 
Due to space constraints, only two representative kernels are randomly selected and shown. 
The target is \texttt{kernel\_duration}, and \texttt{perf\_variation} is computed as 
$\left| \mathrm{avg}(\texttt{kernel\_duration}) - \texttt{kernel\_duration} \right|$. Edge weight defines
change of \% of a feature has positive or negative influence on \texttt{perf\_variation}.}
    \label{fig:all_kernels_combined}
\end{figure}
\vspace{-1.5pt}
\subsection{Kernel Variability}
\subsubsection{Workflow Datasets (Per-Kernel)}: 
In Figure~\ref{fig:per_kerenl_ranks}, we show kernel variability across rank datasets, where each sub-figure presents a causal graph for a per-kernel group defined by its \textit{demangledName}. \texttt{perf\_variability} is computed as 
$\left| \mathrm{avg}(\texttt{kernel\_duration}) - \texttt{kernel\_duration} \right|$.  Black arrows denote positive influence, while red arrows denote negative influence on \texttt{perf\_variablity}. In Figures~\ref{fig:per_kerenl_ranks}a and~\ref{fig:per_kerenl_ranks}b, feature \textit{gridX} shows a positive effect and \textit{gridY} a negative effect, though with differing magnitudes—likely due to architectural or workload factors such as memory access or thread parallelism. For the Rank 1 dataset (in Figure \ref{fig:per_kerenl_ranks}c and \ref{fig:per_kerenl_ranks}d), however, the dominance diagram flips, with the same features exerting opposite influences depending on kernel behavior. Similar shifts appear in Rank 2 and Rank 3 kernels. Overall, these examples illustrate that across all ranks, both the sign and strength of feature-performance relationships vary, showing that kernel-level performance is shaped by rank- and kernel-specific causal dynamics rather than a uniform pattern.

\subsubsection{Workflow Datasets(Entire Dataset)}:
In Figure~\ref{fig:all_kernels_combined}, we extend the analysis with Rank~0 and Rank~3 kernel groups by combining all of the kernel groups. In the Rank~0 causal graph (Fig.~\ref{fig:all_kernels_combined}a), kernel-features such \textit{as gridY} and Memory-related features such as \textit{ dynamicSharedMemory, staticSharedMemory}, and \textit{registersPerThread} contribute positively, whereas \texttt{gridX}, \textit{blockX}, and \textit{blockY} exert weaker negative effects. We also see the same feature influence in the Rank~3 causal graph (Fig ~\ref{fig:all_kernels_combined}b).
% In the Rank~3 graph (Fig.~\ref{fig:rank3_causal}), the same features dominate, though the magnitudes shift: memory effects remain positive, \textit{gridX} stays strongly positive, while \textit{gridY} and \textit{blockX} become more negative. 
% Overall, these cases show that \textit{gridX}, \textit{gridY}, and memory features consistently drive variability, though their strengths vary with rank and workload.

% Figure~\ref{fig:all_kernels_combined} demonstrates the performance variance of all the kernels in a given dataset. The intuition behind this experiment is to discover the overall relationship between parameters and the given target metrics, such as \textit{perf\_variance}. Both figure ~\ref{fig:rank0_causal} and figure ~\ref{fig:rank0_barplot} display the relationships between all the features and the target metrics. From ~\ref{fig:rank0_causal}, we find that \textit{dynamicSharedMemory} , \textit{staticSharedMemory} , and \textit{registeredPerThread} all have black arrows to the \textit{perf\_variation}. We verify the positive correlation with ~\ref{fig:rank0_barplot}, where these features are relatively high across all the time, indicating that when these features fluctuate, they contribute to the increased variance. On top of that, the features having negative effects , \textit{blockX}, \textit{blockY, gridZ} and \textit{localMemoryTotal} , all have red arrows. In the bar plot \ref{fig:rank0_barplot}, their bars are shorter implying less dramatic peaks, suggesting fluctuations in these features don't strongly amplify.

\subsection{Parallel Coordinate of Memory Stall Variability}

\vspace{-0.5pt}
\begin{figure}[!hbtp]
    \includegraphics[width=\linewidth]{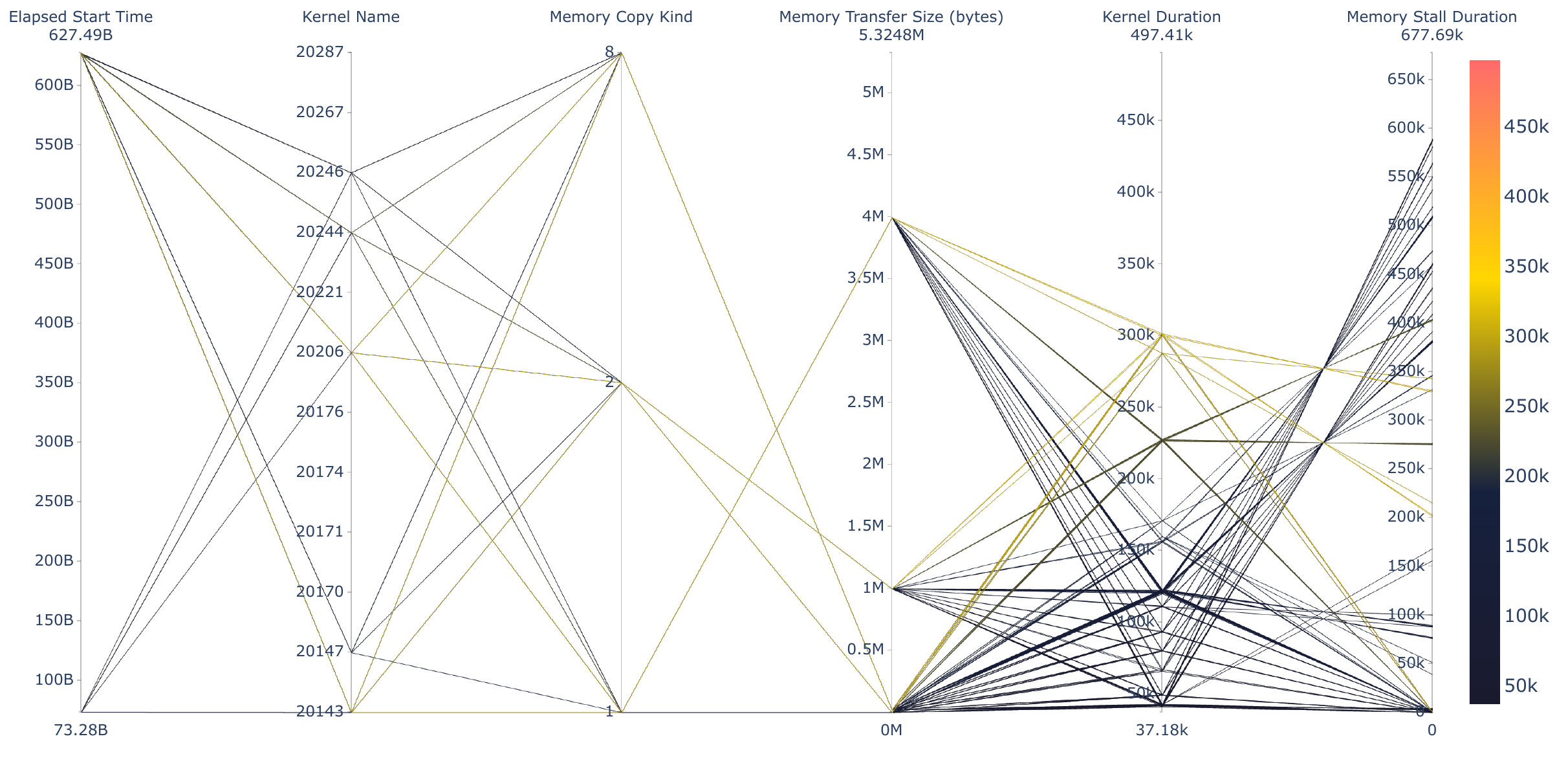}
    \caption{Parallel coordinates plot illustrating kernel execution characteristics under varying memory access scenarios for the Rank 3 dataset. Each line represents an individual kernel invocation, with axes corresponding to elapsed start time, kernel name, memory copy kind, memory transfer size, kernel duration, and memory stall duration. Line colors encode kernel duration, revealing correlations between transfer size and stall time that highlight performance bottlenecks in memory-bound operations.}
    \label{fig:parallel_coordinate}
\end{figure}
\vspace{-3.5pt}

\subsubsection{Impact of Grid Size on Kernel Variability}:
The analysis in Figure~\ref{fig:parallel_coordinate} reveals that a high gridX value is strongly associated with increased kernel variability. Larger grid dimensions imply the launch of a greater number of thread blocks, which, in turn, elevates the complexity of scheduling and load balancing on the GPU. This heightened complexity can introduce fluctuations in execution times across different invocations of the same kernel. Consequently, rather than stabilizing performance, scaling the grid in this way appears to amplify performance dispersion, underscoring the sensitivity of kernel behavior to launch configuration parameters.
\vspace{-0.8em}
\subsubsection{Influence of Copy Kind on Kernel Variability}:
Memory transfer characteristics also show a substantial effect on kernel variability. From Figure~\ref{fig:parallel_coordinate}, we observe operations with copyKind = 8 \textit{(Device-To-Device)} exhibit pronounced instability compared to other transfer modes. This copy kind likely corresponds to a transfer type or direction that imposes additional overhead—such as involving nonstandard paths, asynchronous transfers, or higher latency variability. These irregularities in transfer behavior contribute to inconsistent overlaps with kernel execution, which is reflected in elevated variability metrics.
\vspace{-0.8em}
% \subsubsection{Effect of Source Kind on Kernel Variability}.
% A similar trend is observed with the source type of operations. Kernels associated with srcKind = 2 \ankur{fix this term} demonstrate higher variability than those with other source classifications. This source type may map to a specific memory region or resource with less predictable access characteristics. Such characteristics can lead to irregular kernel start times and latencies, thereby increasing execution-time fluctuations across runs.
% \subsubsection{Independence from Duration Metrics.}
% Interestingly, neither high runtime duration nor high memory duration alone correlates strongly with kernel variability. This finding suggests that longer execution or transfer times do not inherently introduce instability in performance. Instead, kernel variability appears to be driven more by the nature of the configuration and resource types—specifically grid size, copy kind, and source kind—than by the absolute duration of computation or memory operations. This insight emphasizes the importance of examining structural and configuration parameters rather than focusing solely on time-based metrics when seeking to minimize performance variability.

\subsection{PerfCounter Dataset}

Figure~\ref{fig:sub_variability} illustrates the causal relationships within the \texttt{PerfCounter} datasets. All three datasets exhibit identical feature characteristics, and \texttt{gpu\_time} serves as the performance metric, represented as \texttt{perf\_variation\_gpu\_time}. Kernels are categorized by \textit{KERNEL\_NAME}. Consequently, per-kernel and overall causality assessments coincide for \texttt{inversek2j-CUDA} and \texttt{Accuracy-CUDA}. Across all cases, memory-related factors (\texttt{sm\_mem}, \texttt{gpu\_dram}) and instruction metrics \\ (\texttt{sm\_instr}) consistently exhibit a dominant influence on performance variability. In the Accuracy dataset (a), \texttt{gpu\_comp\_mem} and \texttt{avg\_thread\_exec} emerge as significant contributors; whereas, in inversek2j (b), \texttt{sm\_cycles} assumes greater importance. For Kernels 1 and 2, similar causal drivers reappear with differing magnitudes, underscoring workload-dependent causal patterns and stable directional influences with variable intensities across datasets.

\section{Scalability}
% \begin{figure}[!hbtp]
%     \includegraphics[width=0.9\linewidth]{figures/execution_time.pdf}
%     \caption{Scalability of our pipeline. For a fixed input size, we observe that the execution time decreases as the number of MPI ranks increases.}
%     \label{fig:scalability}
% \end{figure}
\begin{figure}[t]
    \centering
    % First subfigure
    \begin{subfigure}[t]{0.48\linewidth}
        \centering
        \includegraphics[width=\linewidth]{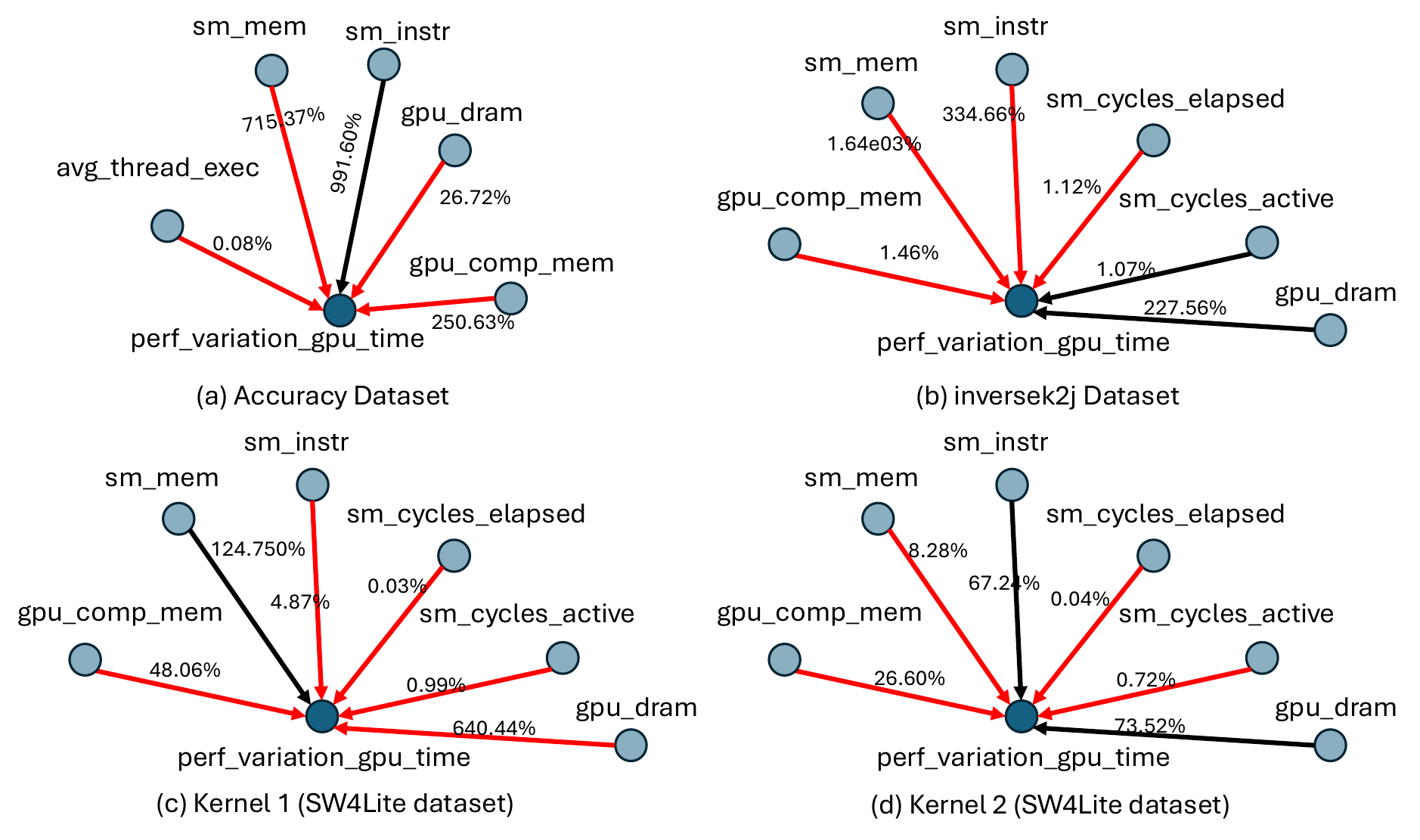}
        \caption{Kernel variability in \texttt{PerfCounter}: black = positive, red = negative influence.}
        \label{fig:sub_variability}
    \end{subfigure}
    \hfill
    % Second subfigure
    \begin{subfigure}[t]{0.48\linewidth}
        \centering
        \includegraphics[width=\linewidth]{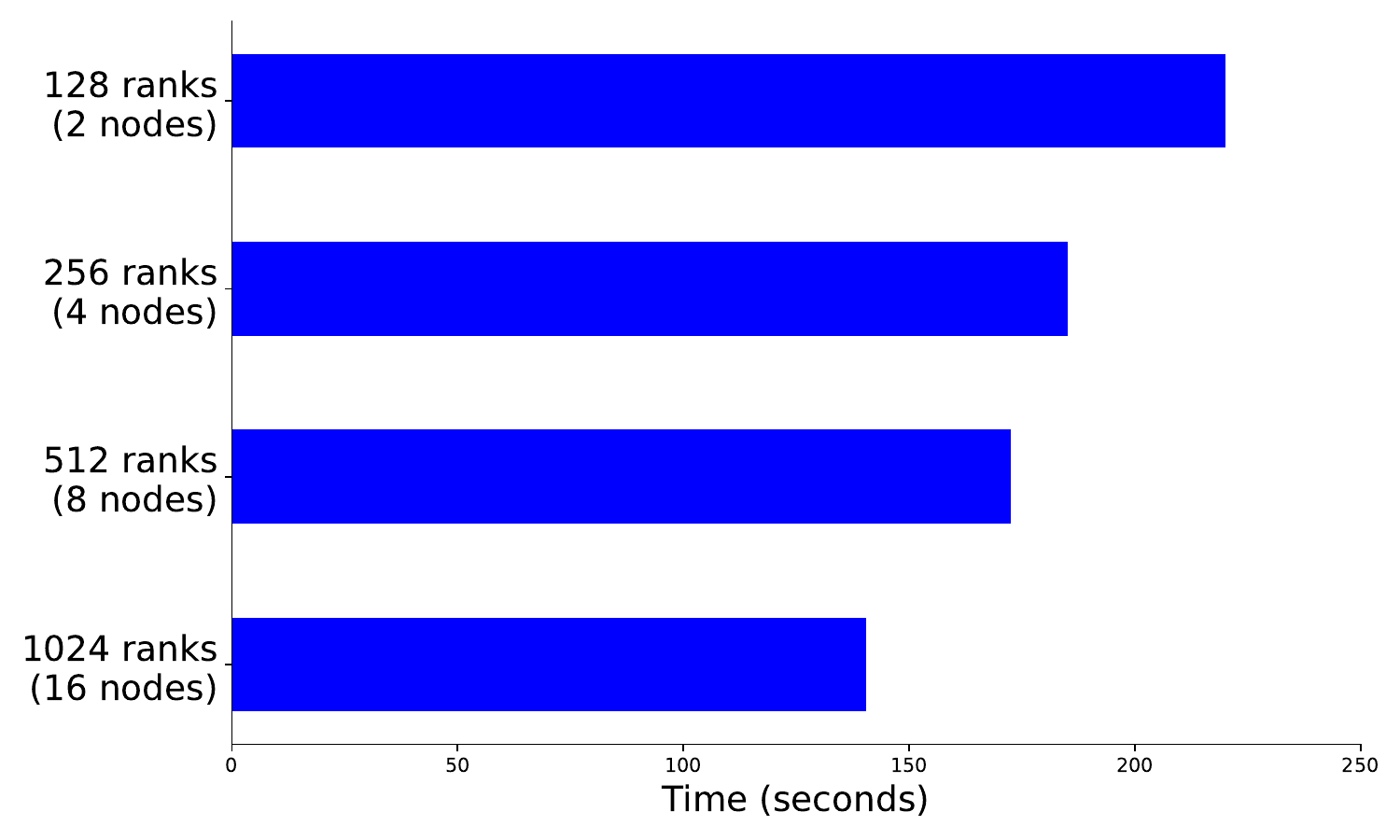}
        \caption{For fixed input, execution time decreases as MPI ranks increase.}
        \label{fig:sub_scalability}
    \end{subfigure}

    \caption{Overview of variability and scalability, illustrating performance behavior across datasets and ranks.}
    \label{fig:combined}
\end{figure}

\vspace{-1em}
% We run memory stall experiments in parallel, consisting of 10.38 minutes of memory traces per workflow dataset. Each dataset contains approximately 93 million samples (Table~\ref{ref:dataset}), and we process four datasets concurrently. With a 10 ms interval, each yields about 68.8K bins, totaling roughly 272.2K bins. From Figure~\ref{fig:sub_scalability}, for a fixed input size, the pipeline shows clear scalability as MPI ranks increase: execution time drops from ~240s at 128 ranks to 180s at 256 (25\%), 120s at 512 (50\%), and 80s at 1024 (67\%). This confirms efficient workload distribution and a steady reduction in runtime with growing concurrency. However, perfect linear speedup is limited by a non-parallelizable setup cost (e.g., input reading, global time range computation, bin partitioning) that does not shrink with more ranks. Communication and synchronization overheads, load imbalance from uneven bins, and I/O contention further reduce efficiency, while aggregation at Rank 0 adds fixed costs, creating a scaling floor.

We run memory stall experiments in parallel, consisting of 10.38 minutes of memory traces per workflow dataset. Each dataset contains approximately 93 million samples (Table~\ref{ref:dataset}), and we process four datasets concurrently. With a 10 ms interval, each yields about 68.8K bins, totaling roughly 272.2K bins. From Figure~\ref{fig:sub_scalability}, for a fixed input size, the pipeline shows clear scalability as MPI ranks increase: execution time drops from ~240s at 128 ranks to 180s at 256 (25\%), 120s at 512 (50\%), and 80s at 1024 (67\%), confirming efficient workload distribution and steady runtime reduction with growing concurrency. However, perfect linear speedup is constrained by non-parallelizable setup operations (e.g., input reading, global time range computation, bin partitioning) and by communication overheads, load imbalance, I/O contention, and Rank 0 aggregation, which collectively impose a scaling floor.

\section{Related Work}
\vspace{-1em}
Modern GPU architectures have become increasingly complex to meet growing demands in scientific computing, AI, and data-driven applications, requiring efficient resource utilization. Heterogeneous systems now feature specialized cores, deeper memory hierarchies, and complex scheduling, making performance bottleneck analysis critical. Yang et al.~\cite{yang2021hierarchical} extended the Empirical Roofline Toolkit to NVIDIA FP64/FP32/FP16 and Tensor Cores, collecting kernel time, FLOPs, and memory traffic with Nsight Compute. Validated on DeepCAM (TensorFlow and PyTorch), their hierarchical Roofline analysis revealed Tensor Core utilization, cache locality, AMP effects, and zero-AI kernel overhead, distinguishing compute- vs. bandwidth-bound behavior. Elvinger et al.~\cite{elvinger2025measuring} profiled fine-grained resource interference (schedulers, IPC/warp, pipelines, caches, memory) on H100 and RTX 3090 using microbenchmarks, exposing metric limits and motivating interference-aware scheduling. For AMD GPUs, Leinhauser et al.~\cite{leinhauser2022metrics} introduced an Instruction Roofline Model using ROCProfiler and BabelStream to enable analysis despite missing counters. They derived IRM equations, built instruction-per-byte and wavefront-normalized GIPS rooflines, and applied them to PIConGPU kernels on V100, MI60, and MI100 for cross-architecture comparison.

In terms of performance modeling in HPC, researchers have also been proposed different idea to find out why performance variability occurs. Wander \cite{lahiry2025wanderexplainabledecisionsupportframework} introduces a comparable technique for causality analysis in the performance domain. However, their study does not provide evidence regarding the method’s scalability or its applicability to large-scale datasets such as ours. Patki et al. ~\cite{patki2019performance} quantify performance variability through repeated runs and CoV, focusing on the extent of fluctuations. PADDLE~\cite{Jay:Paddle} is a unified machine learning framework designed to automate and enhance performance analysis in high-performance computing systems by efficiently handling large-scale performance data and making machine learning more accessible to HPC analysts. Several visualization tools have been introduced in previous studies ~\cite{islam2019toward},~\cite{islam2016a} to effectively illustrate and analyze performance variations in visual data. Though they are visualizing the performance variation, they are still unable to explain the reason "why", which we tried to explain using causality. Opal~\cite{zaeed2025opalmodularframeworkoptimizing} feeds compact, structured performance insights into a template prompt so that an LLM can generate code edits that explicitly tie each change to the observed runtime bottleneck, along with a rationale and deferred suggestions for unsafe edits. However, Our study identifies the causal drivers of performance variance within a given dataset and quantifies their impact, without proposing code-level changes.

\section{Conclusion}
% In this work, we present a distributed causal graph framework for analyzing GPU traces and systematically identifying performance bottlenecks within them. By capturing causal dependencies, our pipeline can provide fine-grained insights into the GPU traces to the stockholders. By pinpointing critical bottlenecks within execution traces, the framework facilitates targeted optimization strategies that can improve GPU utilization and overall system efficiency. A key strength of our pipeline lies in its ability to horizontally scale, allowing it to efficiently handle large and complex input sizes simultaneously. Overall, our findings demonstrate that causal graph analysis is a powerful tool for diagnosing and mitigating GPU performance bottlenecks, providing a rigorous foundation for improving utilization and efficiency across large-scale systems.
\vspace{-1em}
This paper presents a distributed causal graph framework to explain run-to-run variability in GPU execution time. Micro-level fluctuations in kernel behavior often accumulate, leading to significant variability in end-to-end workflow performance. Understanding these fine-grained variations lays the foundation for future hierarchical analyses that attribute the root causes of workflow-level variability. %It scales efficiently to large workloads, offering a rigorous foundation for explaining and mitigating performance variability in large-scale GPU systems.

\section{Acknowledgment}
\vspace{-1em}
This material is based upon work supported by the U.S. Department of Energy, Office of Science under Award Number DE-SC0023173.
This material is based upon work supported by the U.S. Department of Energy (DOE), Office of Science, Office of Advanced Scientific Computing Research, under Contracts DE-AC02-06CH11357.

\bibliographystyle{splncs04}
\bibliography{tzi}

\end{document}